\documentclass[12pt]{spie}  
\usepackage[]{graphicx}
\usepackage{subfig}

\title{Design of the MagAO-X Pyramid Wavefront Sensor}

\author{Lauren H. Schatz\supit{ab}, Jared R. Males\supit{b}, Laird M. Close\supit{b}, Olivier Durney\supit{b}, Olivier Guyon\supit{a}\supit{b}\supit{d}\supit{e}, Michael Hart\supit{a}\supit{c}, Jennifer Lumbres\supit{ab}, Kelsey Miller\supit{ab}, Justin Knight\supit{ab}, Alexander T. Rodack\supit{ab}, Joseph D. Long\supit {b}, Kyle Van Gorkom\supit {ab}, Madison Jean\supit{a}, Maggie Kautz\supit{a}
	\skiplinehalf
	\supit{a} University of Arizona, College of Optical Sciences, 1630 E University Blvd, Tucson, AZ 85719 \\
	\supit{b} University of Arizona, Steward Observatory, Tucson, 933 N Cherry Ave, Tucson, AZ 85721\\
	\supit{c} Institute for Astronomy, University of Hawaii, 34 Ohia Ku St, Pukalani, HI 96768\\
	\supit{d} National Astronomical Observatory of Japan, Subaru Telescope, National Institutes of Natural Sciences, Hilo, HI 96720, USA\\
	\supit{e} Astrobiology Center, National Institutes of Natural Sciences, 2-21-1 Osawa, Mitaka, Tokyo, JAPAN }

\begin{document}
	\maketitle
	
	
	\section{Abstract}
	
	Adaptive optics systems correct atmospheric turbulence in real time. Most adaptive optics systems used routinely correct in the near infrared, at wavelengths greater than 1 $\mu$m. MagAO-X is a new extreme adaptive optics (ExAO) instrument that will offer corrections at visible-to-near-IR wavelengths. MagAO-X will achieve Strehl ratios of $\geq$70$\%$ at H$\alpha$ when running the 2040 actuator deformable mirror at 3.6 $kHz$. A visible pyramid wavefront sensor (PWFS) optimized for sensing at 600-1000 $nm$ wavelengths will provide the high-order wavefront sensing on MagAO-X. We present the optical design and predicted performance of the MagAO-X pyramid wavefront sensor. \cite{PDR}

	\section{Introduction}
	
	MagAO-X is a new visible-to-near-IR extreme adaptive optics system (ExAO) for the 6.5 $m$ Magellan Clay telescope. Working off of lessons learned from MagAO \cite{MagAO} (P.I. Laird Close), and SCExAO \cite{scexao} (P.I. Olivier Guyon), MagAO-X is optimized for high impact science cases, such as a survey of nearby newly formed accreting planets in H$\alpha$. \cite{PDR}. To achieve excellent contrast and resolution. MagAO-X will utilize a 2040 actuator deformable mirror (DM) in conjunction with a cutting-edge coronagraph for starlight suppression. A pyramid wavefront sensor (PWFS) will provide high order wavefront sensing. In this paper we present the design and predicted performance of the MagAO-X pyramid wavefront sensor.
For a general overview of the MagAO-X instrument see Males et al. 2018 \cite{JaredSPIE}, and for the optomechanical design see Close et al. 2018 \cite{LairdSPIE}.

	The pyramid wavefront sensor (1996 by Ragazzoni) \cite{Ragazzoni} is a pupil plane wavefront sensor that in effect works as a 2-D Foucault knife-edge test. In commonly used configurations, the focal plane is split into four quadrants using a double achromatic prism, (LBTAO, and MagAO), or using two roof prisms, (SCExAO). Each quadrant of the focal plane is then reimaged, producing four separate pupil images on the detector. Local wavefront slopes are calculated using the quad-cell centroid of pixel intensities. The number of degrees of freedom controlled by the adaptive optics (AO) system is equal to the number of pixels across one of the pupils in the pyramid wavefront sensor. For our pyramid optic we use a copy of the double achromatic prism used on LBTAO and MagAO. The major design effort is a new camera lens, to image the four pupils onto our OCAM$^2$K detector at the correct size and separation.

	\section{System Design}
	
\begin{figure}[h]
		\centering
		\includegraphics[width=.5\textwidth]{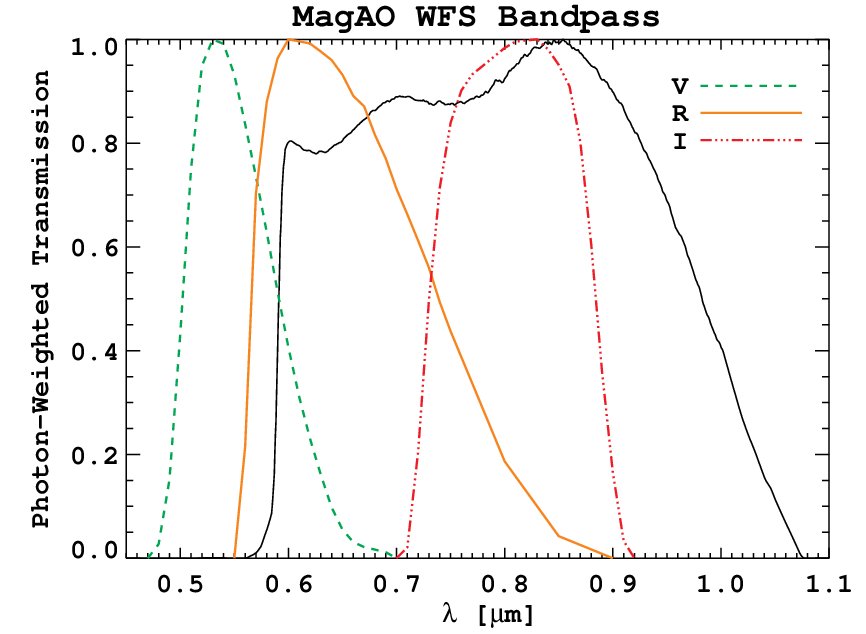}
		\caption{The MagAO pyramid wavefront sensor bandpass, (black curve).\cite{MagAO}}
		\label{fig:bandpass}
	\end{figure}

	The PWFS of the MagAO-X system consists of an achromatic pyramid, a camera lens, and an OCAM$^2$K EMCCD detector. The MagAO-X pyramid wavefront sensor is designed to operate from 600-1000nm bandwidth and a field of view of two arcseconds. Figure \ref{fig:bandpass} is the bandpass of the MagAO PWFS. We expect a similar transmission for the MagAO-X PWFS.

	A new camera lens is designed to meet the requirements of the MagAO-X system. These requirements are listed in Table \ref{tab:requirements}. The OCAM$^2$K will be used in 2x2 binning mode, giving us a 48 $\mu$m pixel size. 
This allows the wavefront sensor to be run at 3.6 kHz. We have chosen a pupil size of 56 pixels across each of the pupils, which is slightly oversampled to prevent aliasing of the higher order modes. The pupil separation of 60 pixels was chosen to maximize the OCAM$^2$K detector. 
	
	\begin{table}
		\begin{center}
			\begin{tabular}{ | l| l | }
				\hline
				Parameter& Requirement \\ \hline
				Wavelength Range &600- 1000 nm \\ \hline
				Pupil Size & 56 pixels; 2.688 mm \\ \hline
				Pupil Separation & 60 pixels; 2.880 mm  \\ \hline
				Pupil Tolerances & $\Delta$ $<$ 1/10th pixel; 2.4 $\mu$m  \\ \hline
				Lens Diameter & 10 mm $<$ D  $<$ 20 mm \\ \hline
				
			\end{tabular}
		\end{center}
		\caption{Parameters for the MagAO-X pyramid wavefront sensor.}
		\label{tab:requirements}
	\end{table}

	\subsection{Pyramid Design}
\begin{figure}[h]
		\centering
		\includegraphics[width=.5\textwidth]{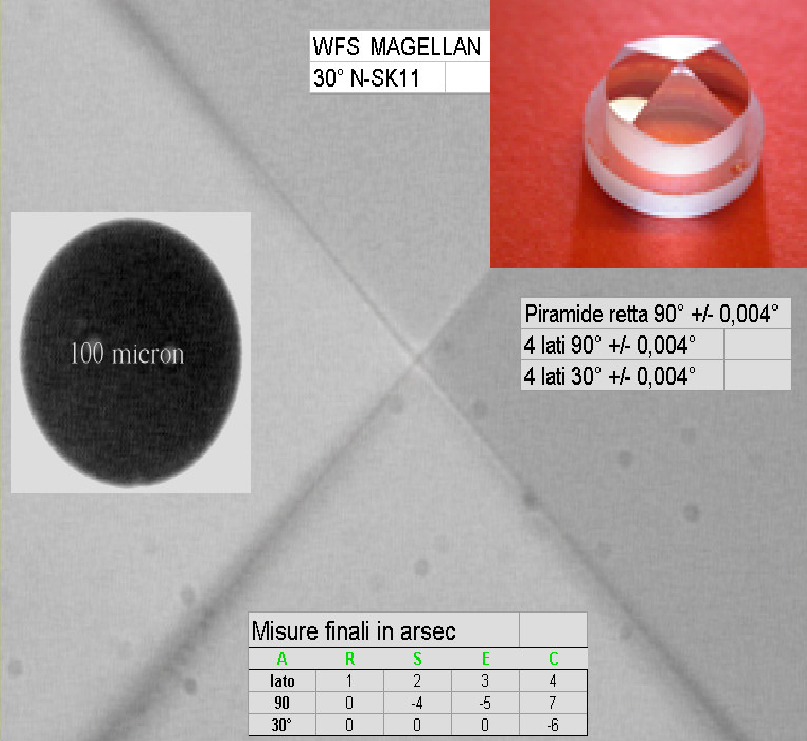}
		\caption{Fabricated pyramid made in Arcetri, Italy by Paolo Stefanini.}
		\label{fig:pyramid}
	\end{figure}

	MagAO-X will be using an excellent achromatic pyramid with a 5 $\mu m$ tip. The pyramid used in the WFS is a double pyramid, consisting of two four sided prisms aligned back to back. Details of the design done by Tozzi et. al. are summarized here.\cite{doubleprism}  A picture of the pyramid is shown in Figure \ref{fig:pyramid}. The total deviation angle needed for the pyramid wavefront sensor is hard to manufacture. Combining two pyramids makes the polishing process easier and at the same time allows us to control chromatic aberrations by using two different glass types. The glass types were chosen using an I.D.L. optimization routine that selected glass combinations from the Shott and Ohara catalog that would give a suitable deflection angle of the double pyramid. The front prism is made from Shott N-SK11, and the back prism is made from Schott N-PSK53.
	
	\subsection{Wavefront Sensor Design}
	
\begin{figure}[h]
		\centering
		\includegraphics[width=.5\textwidth]{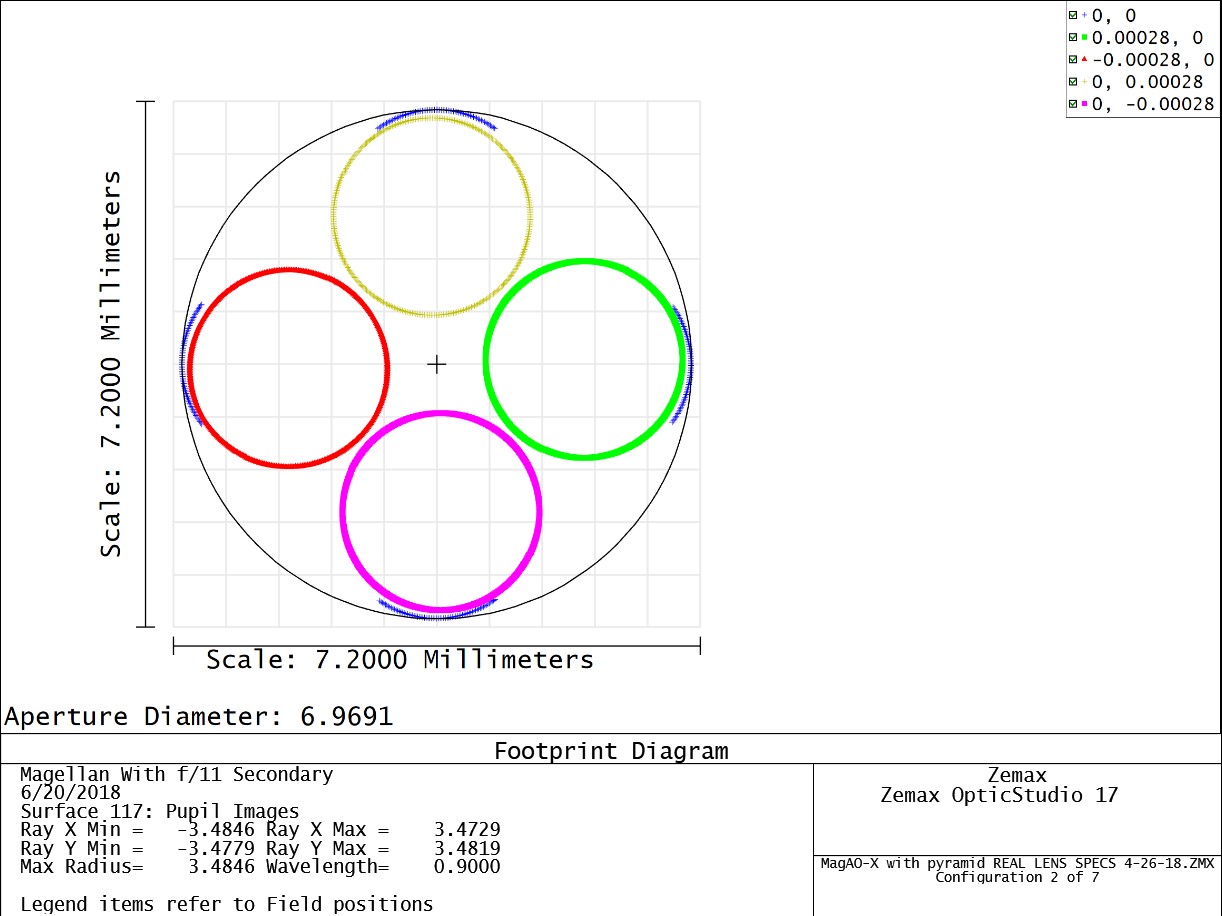}
		\caption{Beam footprint at the image plane.}
		\label{fig:footprint}
	\end{figure}

\begin{figure}%
		\centering
		\subfloat[Optical path in Zemax. ]{{\includegraphics[width=.5\textwidth]{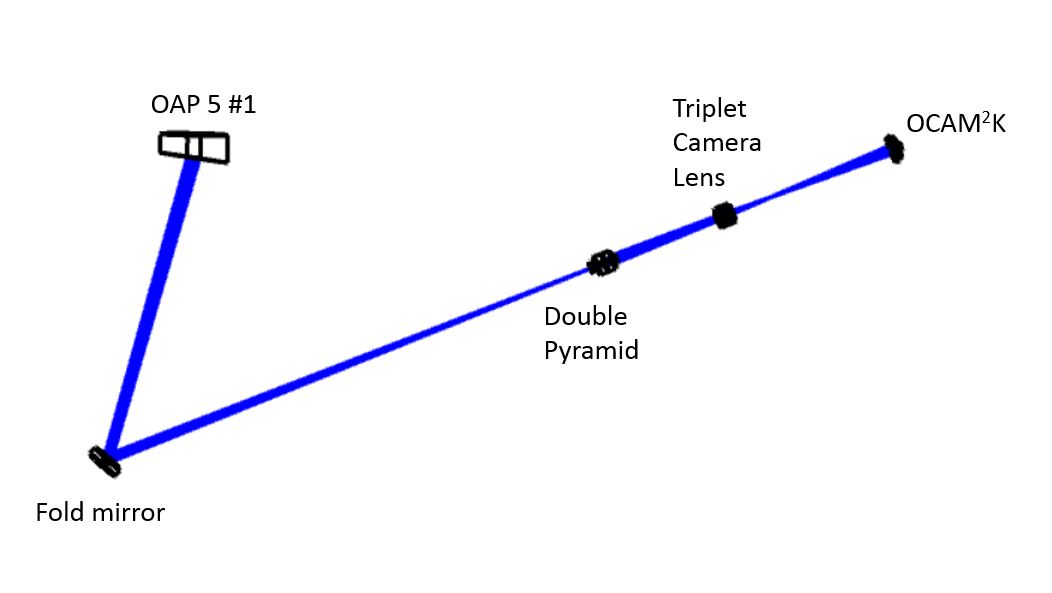} }}%
		\qquad
		\subfloat[Optical path in SolidWorks \cite{LairdSPIE}.]{{\includegraphics[width=.4\textwidth]{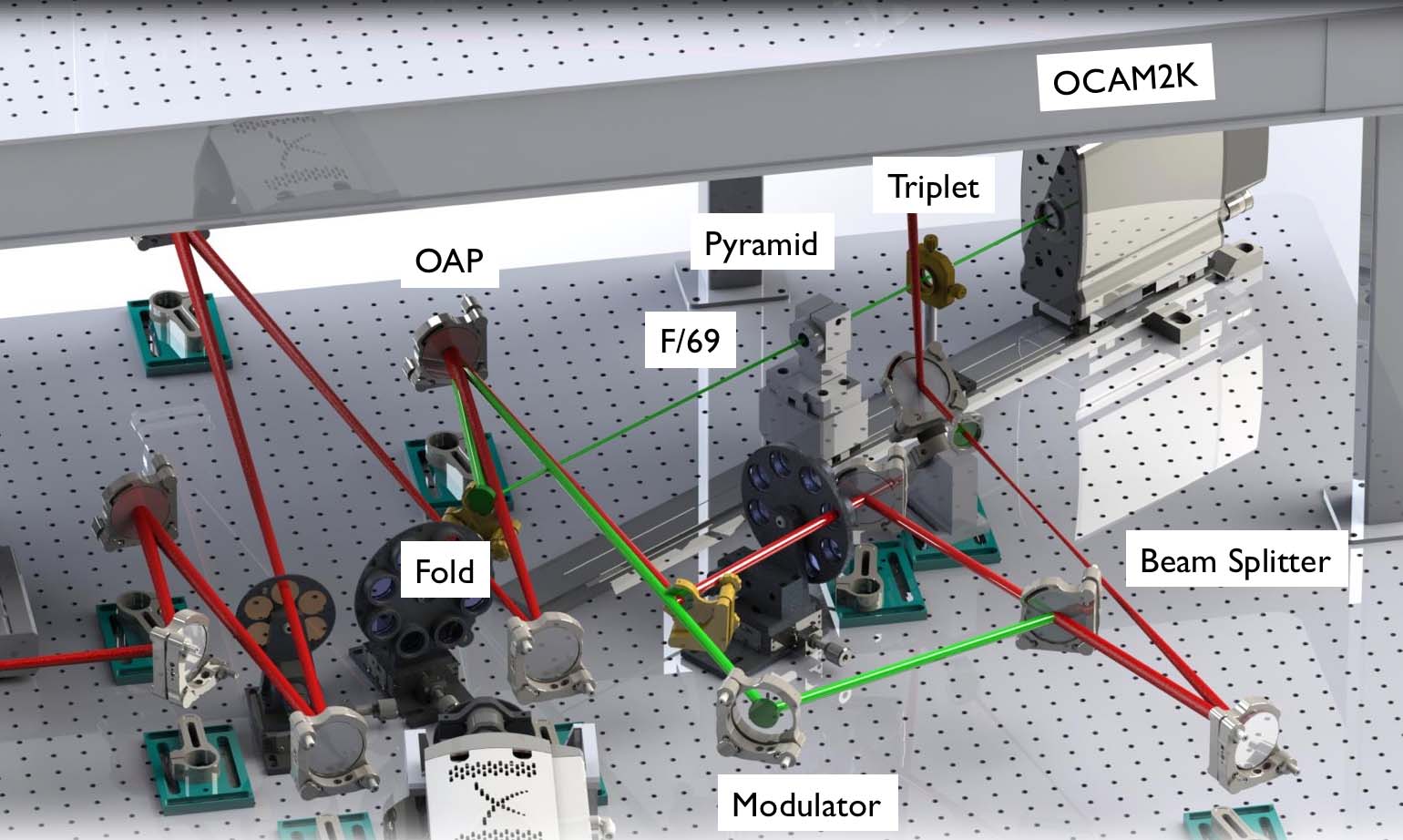} }}%
		\caption{Optical path of the pyramid wavefront sensor. The Zemax ray trace was imported into SolidWorks for the optomechanical design. The red-light path is the science path that goes to the coronagraph. The green light path goes to the pyramid wavefront sensor.}%
		\label{fig:oplayout}%
	\end{figure}

	A design of the wavefront sensor is done in Zemax. An F/69 focus created by an off-axis parabolic mirror is imaged onto the pyramid tip. A custom achromatic triplet then images four pupils onto our OCAM$^2$K wavefront sensor camera. A layout of the wavefront sensor optical path done in both Zemax and SolidWorks is shown in Figure \ref{fig:oplayout}. We reuse the same off axis parabolic mirror seen by the coronagraph arm of MagAO-X. The double pyramid was modeled by the Arcetri team in Zemax, and that same model is used here. A custom achromatic triplet was designed to give the correct pupil size and separation. The two windows in the OCAM$^2$K detector are included in the design for completeness. The expected pupil footprint on the image plane for 800 nm wavelength is given in Figure \ref{fig:footprint}.
	
	\subsection{Achromatic Triplet Design}
\begin{figure}[h]
		\centering
		\includegraphics[width=.5\textwidth]{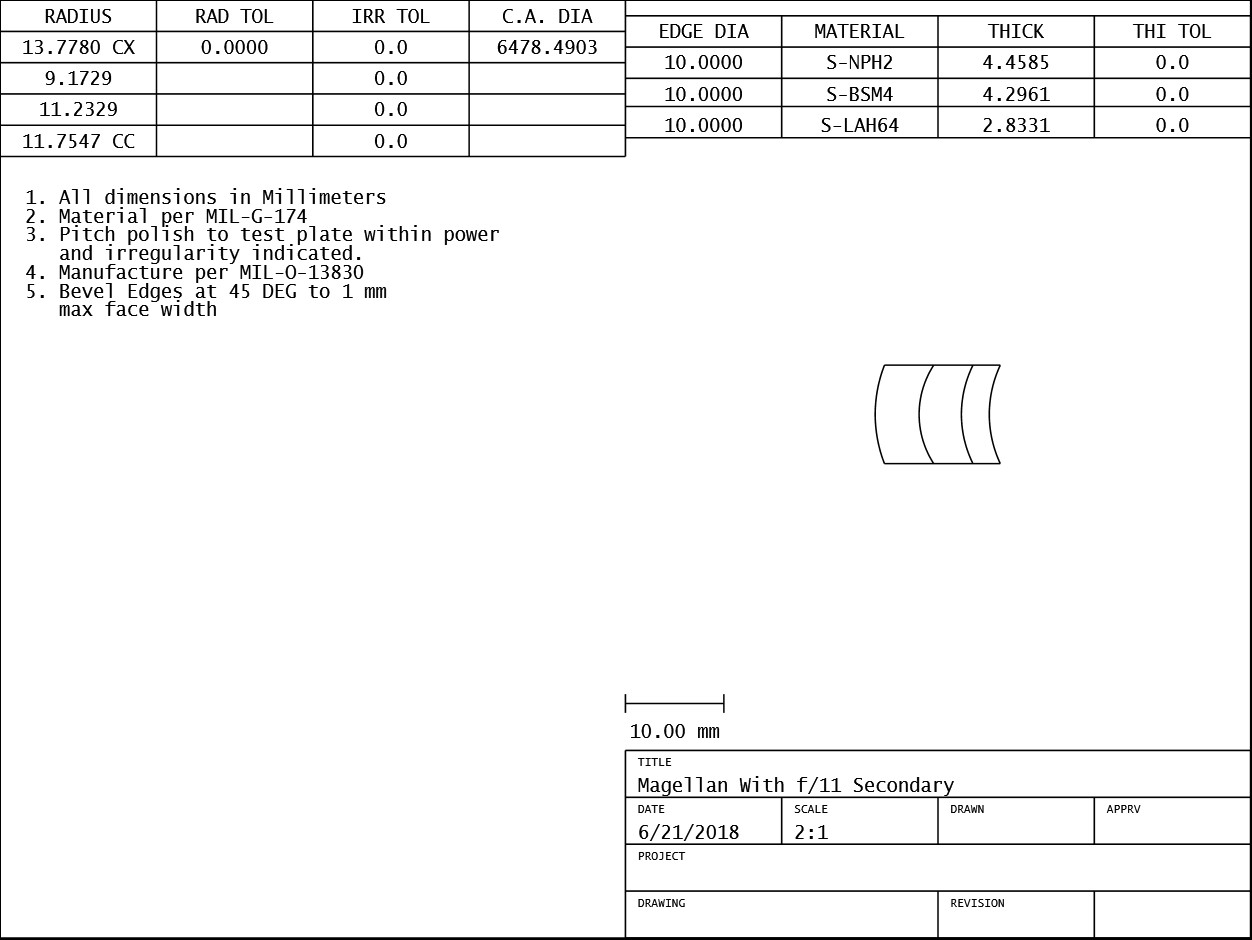}
		\caption{Achromatic triplet.}
		\label{fig:triplet}
	\end{figure}

	The size and separation of the pupils on the detector directly affects the performance of the PWFS. If the pupils are undersized, there will be aliasing in reconstruction of the higher order modes. If the separation between pupils is not correct, complications will arise when pixels are binned on the detector to change the pupil sampling and the integration time.  To ensure the correct size and separations a custom achromatic triplet was designed in Zemax. A schematic of the lens is shown in Figure \ref{fig:triplet}.

\begin{figure}[h]
		\centering
		\includegraphics[width=.5\textwidth]{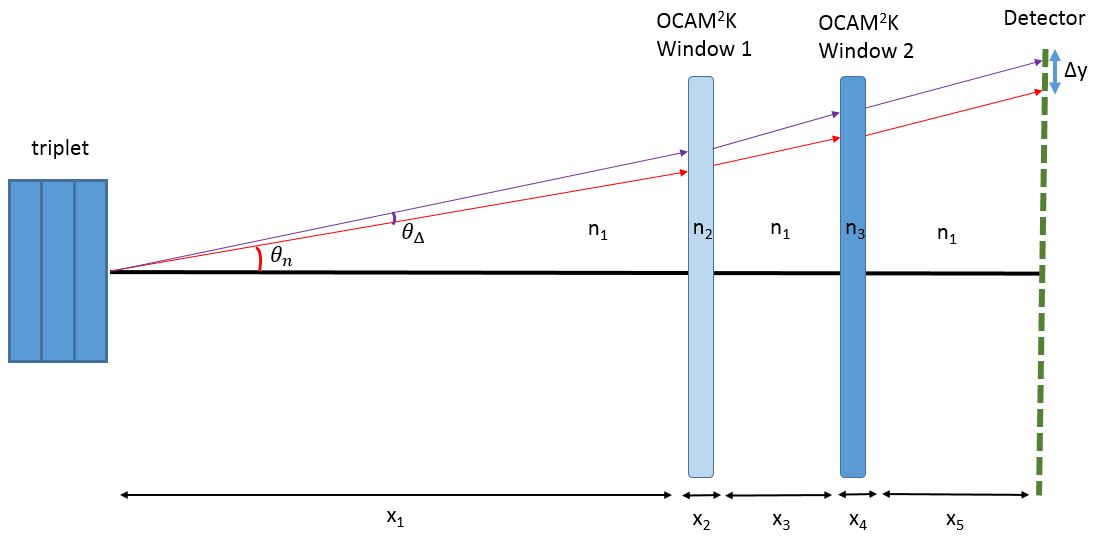}
		\caption{Diagram of the light propagation path used to calculate the change in pupil size.}
		\label{fig:propagation}
	\end{figure}

\begin{figure}[h]
		\centering
		\includegraphics[width=.5\textwidth]{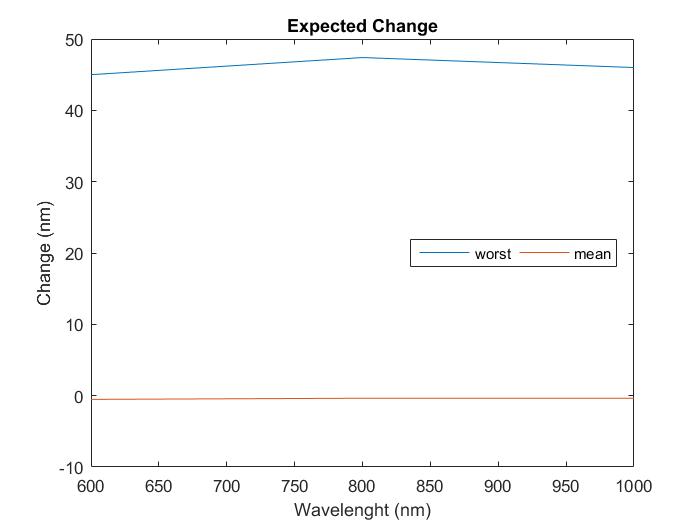}
		\caption{Expected change in pupil size as a function of wavelength.}
		\label{fig:change}
	\end{figure}

	A tolerance analysis to determine lens performance as a function of wavelength is done using parameters from the Precision grade Optimax manufacturing tolerancing chart. Reasonable values of alignment errors were estimated and included in the tolerancing analysis. The figure of merit used was the RMS angular radius of the lens because the pyramid is an afocal system. A 500 trial Monte Carlo simulation was done for three wavelengths, 600nm, 800nm, and 1000nm. At each wavelength the nominal, mean, and worst RMS angular size (twice the angular radius) was recorded. The difference of the mean and worst angles with respect to the nominal value was calculated. That change in angle was propagated through the system to estimate the change in size we would expect. The propagation is shown in Figure \ref{fig:propagation}, where $\theta_n$ is the nominal RMS angular size, and $\theta_\Delta$ is the change in RMS angular size we use to calculate the estimated change $\Delta y$. The distances $x_1 ... x_5$ were taken from the Zemax design, and the indexes $n_1, n_2, n_3$ correspond to air, BK-7, and Sapphire respectively. The index of refraction was adjusted for the different wavelengths when the propagation was calculated using trigonometry and Snell's law. The results are summarized in Figure \ref{fig:change}, where the change in size in nanometers is graphed against wavelength. At worst we expect about a 45 nm change in pupil size and separation, but no change on average. Both are well within our tolerance of the change being no greater than 1/10th a pixel, or 2.4$\mu$m.

	\section{System Performance}

\begin{figure}%
		\centering
		\includegraphics[width=.4\textwidth]{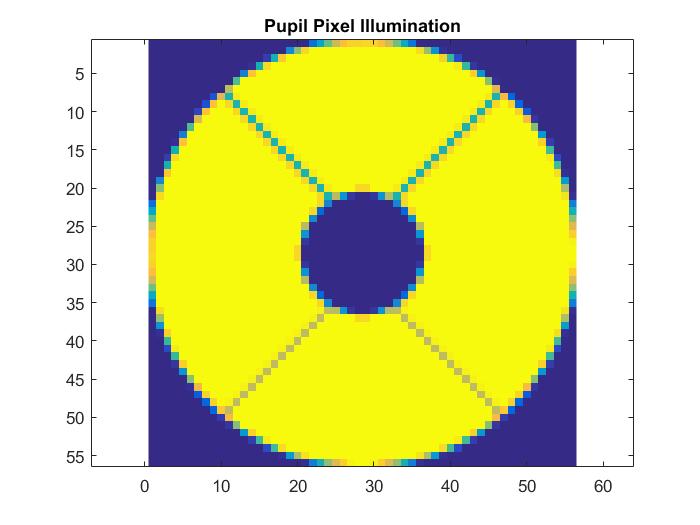}
		\caption{Expected pupil illumination on the PWFS.}	
		\label{fig: pupilpixels}
	\end{figure}

	A simulation of the expected partial illumination of pupil pixels was done in MATLAB. A binary model of the MagAO-X pupil was generated with 10 times the spatial sampling than our expected PWFS pupil. We then bin it down to the expected pupil sampling of our PWFS. That is, we start with a pupil of 560 by 560 pixels, and bin down to a 56 by 56 pixel pupil by summing 10 by 10 pixel bins and normalizing. The expected illumination pattern is shown in Figure \ref{fig: pupilpixels}. A table of the pixel counts is given in Table \ref{tab:actuators}. We expect 1958 fully illuminated pixels within our pupil.

	\begin{table}[h]
		\begin{center}       
			\begin{tabular}{|l|l|} 
				
				\hline
				\rule[-1ex]{0pt}{3.5ex} $\%$ Illumination & $\#$ of Actuators   \\
				
				\hline
				\rule[-1ex]{0pt}{3.5ex} 100$\%$ & 1958  \\
				\hline
				\rule[-1ex]{0pt}{3.5ex} 90$\%$ & 166  \\
				\hline
				\rule[-1ex]{0pt}{3.5ex} 80$\%$ & 24  \\
				\hline
				\rule[-1ex]{0pt}{3.5ex} 0$\%$ & 46  \\
				\hline
				\rule[-1ex]{0pt}{3.5ex} 60$\%$ & 20  \\
				\hline
				\rule[-1ex]{0pt}{3.5ex} 50$\%$ & 18  \\
				\hline
				\rule[-1ex]{0pt}{3.5ex} $<$ 50$\%$ & 904  \\
				\hline
			\end{tabular}
		\end{center}
		\caption{Pixel illuminations in the 56 by 56 pixel pupil. }
		\label{tab:actuators}
	\end{table}
	
	\section{Initial Results}
	
	The camera triplet was manufactured by Rainbow optics and the as-built specifications of the lens was incorporated into the Zemax design. The tolerance of our system was that the pupil size and separation was to be good to 1/10th of a pixel. We found that our fabricated lens was slightly under specifications. The pupils are slightly oversized, and the pupil separation is too close together. We believe this lens can still work for us however, because by iteratively adjusting the positions of the camera lens and camera with respect to each other and the pyramid optic, the sizes and separations of the pupils change. Usually there is a trade off between pupil size, and their seperations for small adjustments in alignment. Meaning, that if you make your pupils smaller, the separation between pupils increases. Table \ref{tab:asbuilt} shows the system requirements of the MagAO-X system, and the expected performance with our fabricated lens.

	\begin{table}
		\begin{center}
			\begin{tabular}{ | l| l |  l |}
				\hline
				Parameter& Requirement & As Built\\ \hline
				Wavelength Range &600- 1000 nm& 600-100 nm\\ \hline
				Pupil Size & 56 pixels; 2.688 mm& 2.696 mm\\ \hline
				Pupil Separation & 60 pixels; 2.880 mm& 2.857 mm \\ \hline
				Pupil Tolerances & $\Delta$ $<$ 1/10th pixel; 2.4 $\mu$m& $\Delta_{size}$=8 $\mu m$,  $\Delta_{sep}$=-23 $\mu m$ \\ \hline
				Lens Diameter & 10 mm $<$ D  $<$ 20 mm & D= 10.1 mm\\ \hline
				
			\end{tabular}
		\end{center}
		\caption{Parameters for the MagAO-X pyramid wavefront sensor and the as built expected performance from our Zemax model.}
		\label{tab:asbuilt}
	\end{table}
	
\begin{figure}%
		\centering
		\subfloat[Mounted triplet lens. ]{{\includegraphics[width=.3\textwidth]{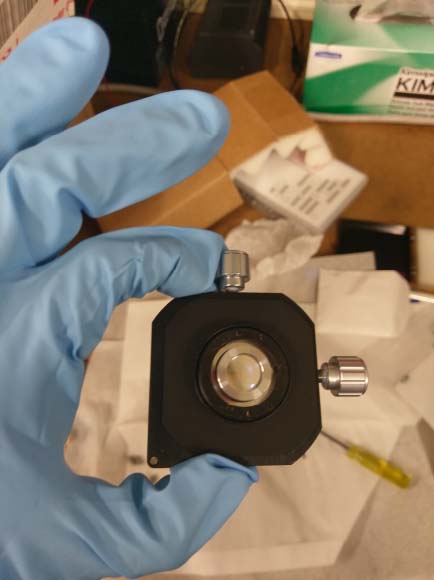} }}%
		\qquad
		\subfloat[Aligned pyramid wavefront sensor]{{\includegraphics[width=.5\textwidth]{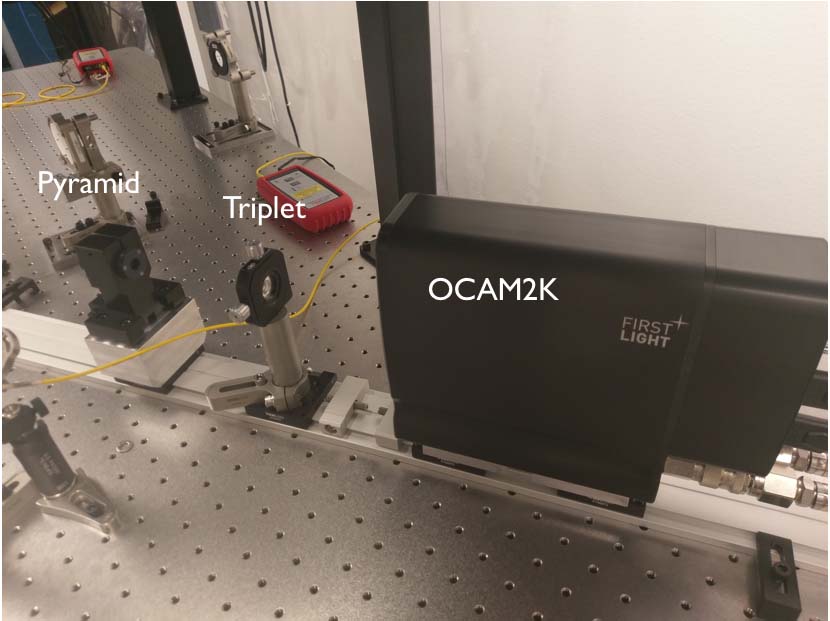} }}%
		\caption{ The MagAO-X pyramid wavefront sensor initial alignment.}%
		\label{fig:mounts}%
	\end{figure}

\begin{figure}%
		\centering
		\subfloat[Pupils from HeNe light source. ]{{\includegraphics[width=.4\textwidth]{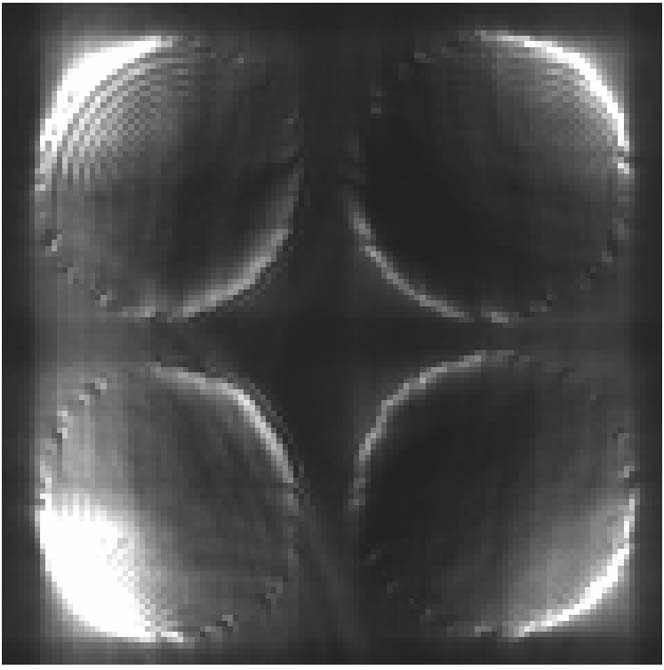} }}%
		\qquad
		\subfloat[Pupils from white light source.]{{\includegraphics[width=.41\textwidth]{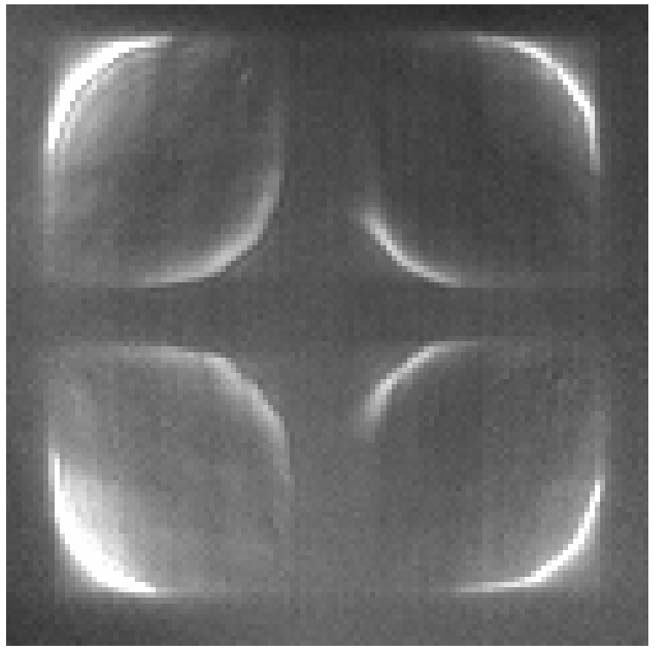} }}%
		\caption{ The initial pyramid wavefront sensor pupils on the OCAM$^2$K detector.}%
		\label{fig:pupils}%
	\end{figure}

	An initial alignment of the MagAO-X pyramid wavefront sensor was done using a HeNe single mode fiber laser, two off axis parabolic mirrors, and a temporary pupil mask with coarse edges. The pyramid was not modulated during alignment. The pyramid optic, camera lens, and OCAM$^2$K were mounted on a coaxial rail system from ThorLabs. Custom mounting plates were fabricated for each optic, so that each could be mounted onto rail carriages and meet the beam height requirement of 5 inches. Figure \ref{fig:mounts} shows the mounted camera lens, and the initial alignment of the pyramid wavefront sensor. Images of the unmodulated pupils are shown in Figure \ref{fig:pupils}. Figure \ref{fig:pupils} (a) shows the pupils illuminated by a HeNe fiber laser. The edge of the pupil is unclean due to the coarse edges of our temporary pupil. Figure \ref{fig:pupils} (b) shows the pupils illuminated with a white light source. An iris was put in front of the fake pupil in the system to clean up the edges of the pupil.

	\section{Conclusions and Future Work}
	The MagAO-X pyramid wavefront sensor has been optimized for wavefront sensing from 600 to 1000 $nm$. A custom acromatic triplet was designed in Zemax to give the appropriate pupil seperation (60 pixels center to center) and sampling (56 pixels across the pupil). We expect to control a maximum of 1958 modes on our 2048 actuator deformable mirror. Future work will see the pyramid wavefront sensor comissioned. In parallel we will be working with the University of Arizona Wavefront Control testbed to study the performance of different pyramid architectures and reconstruction matrices. 
	
	\pagebreak

\end{document}